**ARTICLE**                                                                                    **Open Access**

# Enhancing sensitivity in atomic force microscopy for planar tip-on-chip probes

H. Tunç Çiftçi[1], Michael Verhage[1], Tamar Cromwijk[1], Laurent Pham Van[2], Bert Koopmans[1], Kees Flipse[1] and Oleg Kurnosikov[1,3 ✉]

## Abstract
We present a new approach to tuning-fork-based atomic force microscopy for utilizing advanced "tip-on-chip" probes with high sensitivity and broad compatibility. Usually, such chip-like probes with a size reaching $2 \times 2$ mm$^2$ drastically perturb the oscillation of the tuning fork, resulting in poor performance in its intrinsic force sensing. Therefore, restoring initial oscillatory characteristics is necessary for regaining high sensitivity. To this end, we developed a new approach consisting of three basic steps: tuning-fork rebalancing, revamping holder-sensor fixation, and electrode reconfiguration. Mass rebalancing allows the tuning fork to recover the frequency and regain high Q-factor values up to $10^4$ in air and up to $4 \times 10^4$ in ultra-high vacuum conditions. The floating-like holder-fixation using soft wires significantly reduces energy dissipation from the mounting elements. Combined with the soft wires, reconfigured electrodes provide electrical access to the chip-like probe without intervening in the force-sensing signal. Finally, our easy-to-implement approach allows converting the atomic force microscopy tip from a passive tool to a dedicated microdevice with extended functionality.

## Introduction

In the scope of ever-expanding fundamental studies and research in nanotechnology, the scanning probe microscope (SPM) has shown a permanent improvement as the ultimate tool for locally probing interactions. Shortly after the debut of scanning tunneling microscopy (STM)[1] and atomic force microscopy (AFM)[2] in the 80s, several variations started branching off to a broad range from non-invasive near-field detection[3] to probe-assisted nanolithography[4]. This development was accompanied by a transition from soft silicon cantilevers with tiny tips (Fig. 1a–c) toward robust tuning-fork-based force sensors with tip-on-chip probes (Fig. 1d). In this paper, we will demonstrate a new approach to tuning-fork-based AFM to utilize chip-like oversized probes with enhanced sensitivity and tip functionalization (Fig. 1e).

Early variations aimed at application-specific SPM just by coating a traditional Si-tip, as shown in Fig. 1b. For instance, magnetically coated tips paved the way for spin-polarization sensitive STM (SP-STM)[5] and additional magnetic stray-field detection in AFM (MFM)[6–10]. Originating from the same principle, superconductive STM-tips[11,12], conductive atomic force microscopy (CAFM)[13], and Kelvin probe force microscopy (KPFM)[14,15] extended the scope of first-generation tip metallization. Further developments in the probe metallization converted the passive tip into an active element with current control, as illustrated in Fig. 1c. Especially self-heating Millipede-tips[16,17] and scanning thermal microscope (SThM) with integrated Wollaston probe[18] already highlighted the potential of electrical access to the tip end in AFM applications.

Although such functionalized probes provided unique solutions, their applicabilities remained limited in specific operations. The reason was usually either the soft silicon cantilever's fragility or the tiny needle-like tip. Firstly, it is well known that conventional soft Si-cantilevers and their optical beam-based feedback systems can be limiting in

Correspondence: Oleg Kurnosikov (oleg.kurnosikov@univ-lorraine.fr)
[1]Department of Applied Physics, Eindhoven University of Technology, PO Box 513,5600 MB Eindhoven, the Netherlands
[2]DRF/IRAMIS/SPEC-LEPO, Centre CEA de Saclay, 91191 Gif-sur-Yvette, France
[3]Institut Jean Lamour, Lorraine University, 54000 Nancy, France

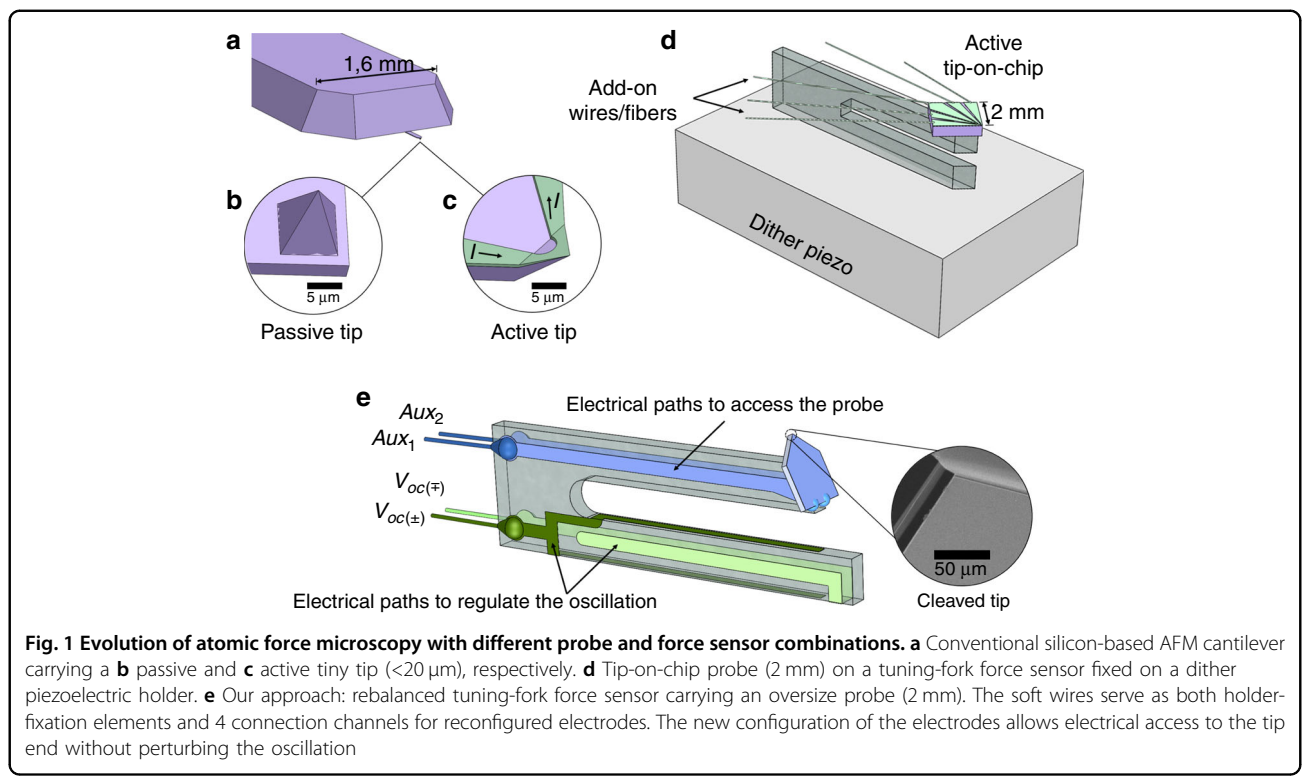

**Fig. 1 Evolution of atomic force microscopy with different probe and force sensor combinations. a** Conventional silicon-based AFM cantilever carrying a **b** passive and **c** active tiny tip (<20 μm), respectively. **d** Tip-on-chip probe (2 mm) on a tuning-fork force sensor fixed on a dither piezoelectric holder. **e** Our approach: rebalanced tuning-fork force sensor carrying an oversize probe (2 mm). The soft wires serve as both holder-fixation elements and 4 connection channels for reconfigured electrodes. The new configuration of the electrodes allows electrical access to the tip end without perturbing the oscillation

vacuum environments, high/low temperatures, or light-sensitive material investigations. Secondly, the geometric entities of tiny tips are highly challenging. It is inherently neither easy to grow a uniform film on a tiny tip nor to precisely characterize the film's growth on a conical or pyramidal tip form. As a result, the unknown magnetic stray-field distribution at the needle-like tip end causes errors in extracting meaningful quantitative value[19]. Thereby, one could suggest micro-tailored structures similar to the SThM[18] as a solution for quantitative sensing and further advanced tip functionalization. However, their feasibility as a scanning tip might be considered questionable due to the fabrication costs and poor durability of such fragile microelectronic systems.

An alternative route beyond traditional needle-like SPM-tips is via probes in planar form. A planar probe is simply a wafer with a sharp corner serving as a scanning tip. Siahaan et al.[20] used a metallic film-covered glass slab as an STM-tip that demonstrated an atomically resolved topography of single-crystalline copper. In addition to the high resolution, a planar probe provides unmatched flexibility in combining ultra-thin film technology and nano-tailoring near its sharp tip. To this end, Leeuwenhoek et al.[21] developed a double-tip STM with a reduced interprobe distance by tailoring a chip-like structure over a cleavable thin-film probe. Such freedom in ultra-thin film engineering near the tip end is far beyond the limits of what traditional needle-like configurations can reach.

However, such an advanced tip-on-chip implementation is far more complex for AFM than for STM. Conventional silicon cantilevers are not suitable for chip-like probes due to their limited load-carrying capacity and their narrow scope of operational conditions. Therefore, the advent of quartz-tuning-fork-based detection systems was a milestone for extending AFM studies[3,22–26]. In addition to compatibility with the vacuum environment and high/low temperatures, stiff quartz-made tuning-fork oscillators can also carry heavier probes. Among their configurations demonstrated so far, we can make a distinction between single-prong and two-free-prong oscillators, as explained below.

The sensing capability collapses in a standard tuning fork when the attached probe yields a severe asymmetry between two prongs[27]. Evading the symmetry-dependent sensitivity of the two-free-prong systems, Giessibl et al.[28,29] introduced a single-prong type qPlus approach by immobilizing one tine, which revealed an improved load-carrying capacity. In this qPlus configuration as illustrated in Fig. 1b, one prong remains fixed to a dither piezoelectric block while the other tine oscillates an oversize probe enough for non-contact regime imaging in UHV.

As of the early 2000s, this single-prong type oscillator led to state-of-the-art prototypes of planar probe utilization in AFM studies. For instance, scanning SQUID force microscopy (SSFM)[30], scanning Hall probe microscopy (SHPM)[31], and near-field scanning microwave microscopy (NFSMM)[32,33] propelled the chip-like probe approach into a thriving platform for advanced AFM applications. However, despite its popularity among the unorthodox probe approaches, using a chip-like probe instead of a tiny needle puts a lower ceiling on the Q-factor (Q) value of a qPlus system.

As Giessibl presented an extensive comparison in their work[29], combining the most recent amplifier and a qPlus sensor carrying a tiny tip can have sufficiently high Q values nearing $3 \times 10^3$ in air and up to $15 \times 10^3$ in UHV at room temperature. Depending on the added probe's mass, a qPlus system with a tip-on-chip instead of a tiny needle usually has a Q value of around a few hundred and a resonance frequency typically between 15 and 30 kHz[30–33]. However, a standard AB38T model or a similar quartz-tuning fork with two-free-prongs used for fabricating qPlus originally has a Q value on the order of $10^4$ at its resonance frequency of 32.768 kHz under ambient conditions. In other words, the loss in qPlus sensitivity is considerably high when carrying a chip-like probe. At this point, we focused on one critical question: Is it possible to restore initial oscillatory characteristics and regain high sensitivity after attaching an oversized (e.g., tip-on-chip) probe without immobilizing one prong?

In 2014, Ooe et al.[34] demonstrated a systematic work on how the dissipation varies with frequency shift differently in a single-prong type qPlus and a two-free-prong oscillator. Their study goes beyond a two-system comparison by emphasizing oscillation recovery possibilities of a two-free-prong oscillator by the "retuning" method[34]. This retuning method as a mass compensation between two prongs after attaching a probe is not a brand new concept[35,36]. However, the retuning method of Ooe et al.

steps forward by recovering the oscillation characteristics near its initial values, revealing a Q value of $1.8 \times 10^4$ at its resonance frequency of 32.877 kHz while carrying a probe. In this scope, we believe that the "retuned" tuning forks in the two-free-prong configuration method can be a breakthrough for utilizing oversized chip-like probes.

In recent work[37], we introduced a proof of Ooe's retuning concept[34] in utilizing chip-like planar probes with a more reliable and efficient method of rebalancing, as illustrated in Fig. 1e. In the present paper we demonstrate a greatly improved fabrication procedure providing a versatile system with broad compatibility, and a high sensitivity providing atomic resolution at room temperature in both AFM and STM modes. More specifically, we propose a new tuning-fork design based on three pillars: tuning-fork rebalancing, revamping holder-sensor fixation, and electrode reconfiguration for electrical access to the chip-like planar probes.

### Fabrication

Here we present the rebalancing method[34] to regain high Q values for utilizing heavy tip-on-chip probes on commercially available AB38T quartz-tuning forks. A default AB38T oscillator has two identical prongs ($\Delta m = 0$) surrounded by electrodes to track and control the oscillation via two channels, as given in Fig. 2a. We start by polishing away some load from one prong. As a result, we obtain two asymmetric prongs for which the mass difference can be expressed as $\Delta m = m_2 - m_1 > 0$ as given in Fig. 2b. Note that this difference should be approximate to the extra mass from probe attachment ($\Delta m \approx m_{em}$). At this point, we apply one more step prior to tip attachment, unlike the previous tip-on-chip approaches.

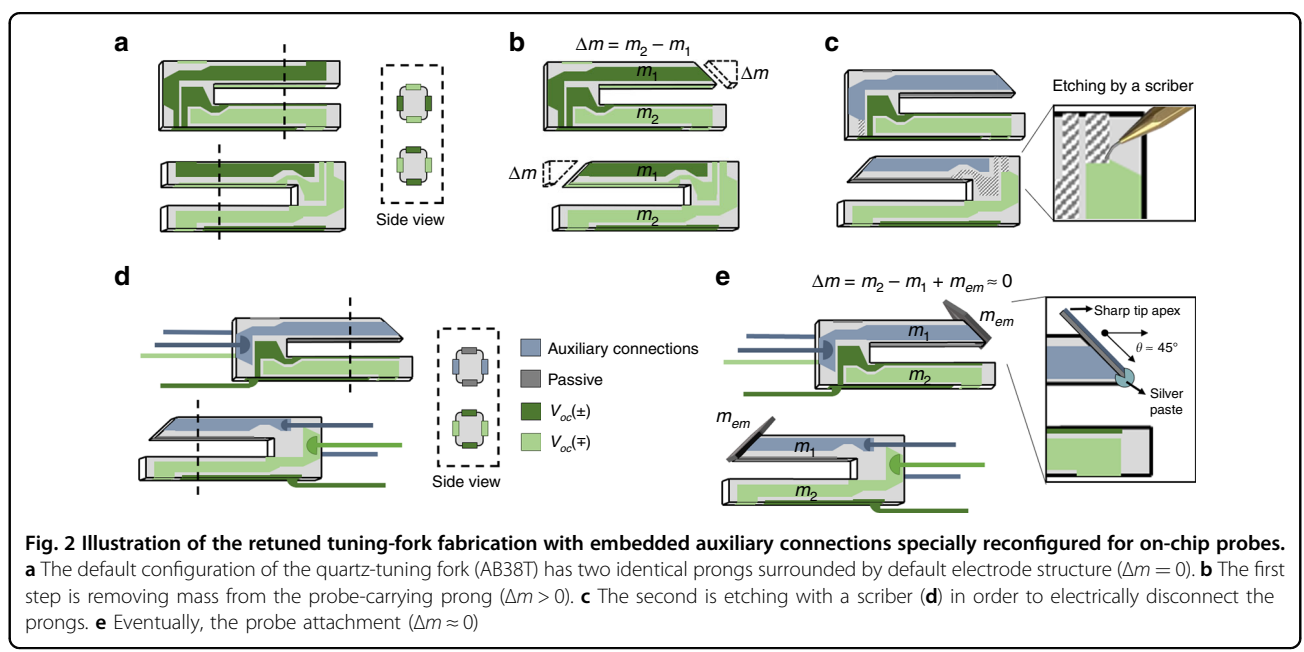

**Fig. 2 Illustration of the retuned tuning-fork fabrication with embedded auxiliary connections specially reconfigured for on-chip probes.** **a** The default configuration of the quartz-tuning fork (AB38T) has two identical prongs surrounded by default electrode structure ($\Delta m = 0$). **b** The first step is removing mass from the probe-carrying prong ($\Delta m > 0$). **c** The second is etching with a scriber (**d**) in order to electrically disconnect the prongs. **e** Eventually, the probe attachment ($\Delta m \approx 0$)

The mass balancing alone is not enough since a sophisticated chip-like planar probe requires auxiliary connections. Therefore, we manipulate existing electrodes as embedded auxiliary connections to eliminate potential dissipation from traditionally applied add-on wires. Figure 2c shows how we etch the electrodes locally with a scriber to disconnect the prongs electrically. One might question the effect of this manipulation on the signal readout capability. However, as long as we use a piezoelectric actuator for scanning, one of the mechanically coupled prongs is enough to trace the oscillation signal. At the same time, the other 'retuned' prong serves as the probe-carrying cantilever. As Fig. 2d shows, the reconfigured electrical paths on the probe-carrying prong provide links between auxiliary connections and the chip-like probe.

Following the mass removal and the electrode reconfiguration, the next step is setting the probe, as illustrated in Fig. 2e. The probe attachment by gluing on the mass reduced prong rebalances the effective mass on the prongs as $\Delta m = m_2 - m_1 + m_{em} \approx 0$.

Tip attachment is preferably at an ~45° angle to utilize the atomically sharp cleaved tip end better. The SEM image in Fig. 1e demonstrates how the angled probe fixation results in imitating the pyramid-like traditional tip geometry, as given in Fig. 1b. As much as the retuning and the electrode reconfiguration methods, we should also consider the holder-fixation method. To this end, we used soft wires that can reduce damping and connect the electrodes via four channels, as given in Fig. 2d. Another advantage of this floating-like holder-fixation is high compatibility with various systems without losing scan performance, as revealed by the experimental results in the following section.

## Results

To demonstrate the capabilities of our new approach in terms of high sensitivity and broad compatibility, we performed several scans in AFM and STM modes over a diverse range of samples from crystals to polymers. For the measurements, we used two different setups operating in air and ultra-high vacuum (UHV), respectively. Besides its versatility, easy adaptation to two different systems provides an insight into the compatibility of the new approach.

The first system for scanning under ambient conditions is a modified NT-MDT Spectra SPM system, as shown in Fig. 3. The control unit regulates the oscillation via a USB mini-A type adaptor (Fig. 3a) between the AFM head (Fig. 3b) and the force sensor (Fig. 3c). The system regulates the oscillation at a constant frequency mode in the range of 0.5–10 nm. For the UHV measurements, we adapted the planar probe-carrying tuning-fork sensor for a Scienta Omicron Variable Temperature (VT) SPM system in a UHV chamber with a $10^{-10}$ mBar base pressure as given in Fig. 4a-c. The UHV system is also operated in the constant frequency mode of

AFM, regulating the oscillation amplitude between 0.1 and 1 nm. For STM operations, we used constant height mode.

Figure 3d-f demonstrates air AFM scans using the NT-MDT Spectra SPM system. The AFM scan in Fig. 3d shows the topography of the surface of an Ir/Pt/Co multilayered thin film with a resolution reaching down to a few tens of nanometers, as the plot in Fig. 3g reveals. As another example, Fig. 3e demonstrates the scan capabilities of the bulky tip apex reaching 250 nm deep in the large grating patterns of a silicon calibration sample, as shown by the plot in Fig. 3h. Other than these stable surfaces, the planar probe was also successfully used to scan surfaces that are deformable under small forces, such as the topography of paraffin wax in the air shown by Fig. 3f with a surface roughness of a few nanometers, as given by Fig. 3i.

Examples of using the planar probe in the UHV system are given in Fig. 4c-f. The first specimen studied in UHV is epitaxial graphene (EG) grown by thermal decomposition on silicon carbide (SiC). We imaged the SiC graphene in combined STM and AFM mode in a repulsive force regime due to the deformation of graphene[38]. As shown in Fig. 4c-f, both AFM and STM mode scans over this sample demonstrated the atomic honeycomb structure of graphene and the well-known $(6\sqrt{3} \times 6\sqrt{3})R30°$ hexagonal periodic arrangement[38,39].

These results provide unambiguous evidence for atomic resolution using planar probes not only in STM, but even in AFM mode.

Such achieved high resolution points out that the cleaved tip radius at the end of an oversized chip-like probe is sufficiently small on the scale of several nanometers. At the same time, the raster scans in and out of high aspect grating patterns of the Si calibration sample (Fig. 3e) profiled the tip convolution (Fig. 3h) of the planar probe with a large opening angle. Such high stability and versatility emanate from the tuning fork's robustness, which prevents the tip from jump-to-contact behavior even on easy-to-deform surfaces (Fig. 3f). As a result, the control unit could perform reliable surface scans in many consecutive measurements both in the constant frequency mode and the constant height mode. The prolonged scans taken in the repulsive force regime on the SiC-graphene sample without a noticeable tip degradation emphasized the cleaved tips' mechanical durability (Fig. 4).

## Discussion

This section discusses quantitatively how the retuned tuning forks with the tip-on-chip can reveal a remarkable scan performance consistently in different systems. To this end, we will break down the multilayered fabrication process to unfold the purpose of each step. Before that, however, it is essential to start from fundamental parameters affecting AFM sensitivity to understand the choices we made in developing this new force-sensing approach.

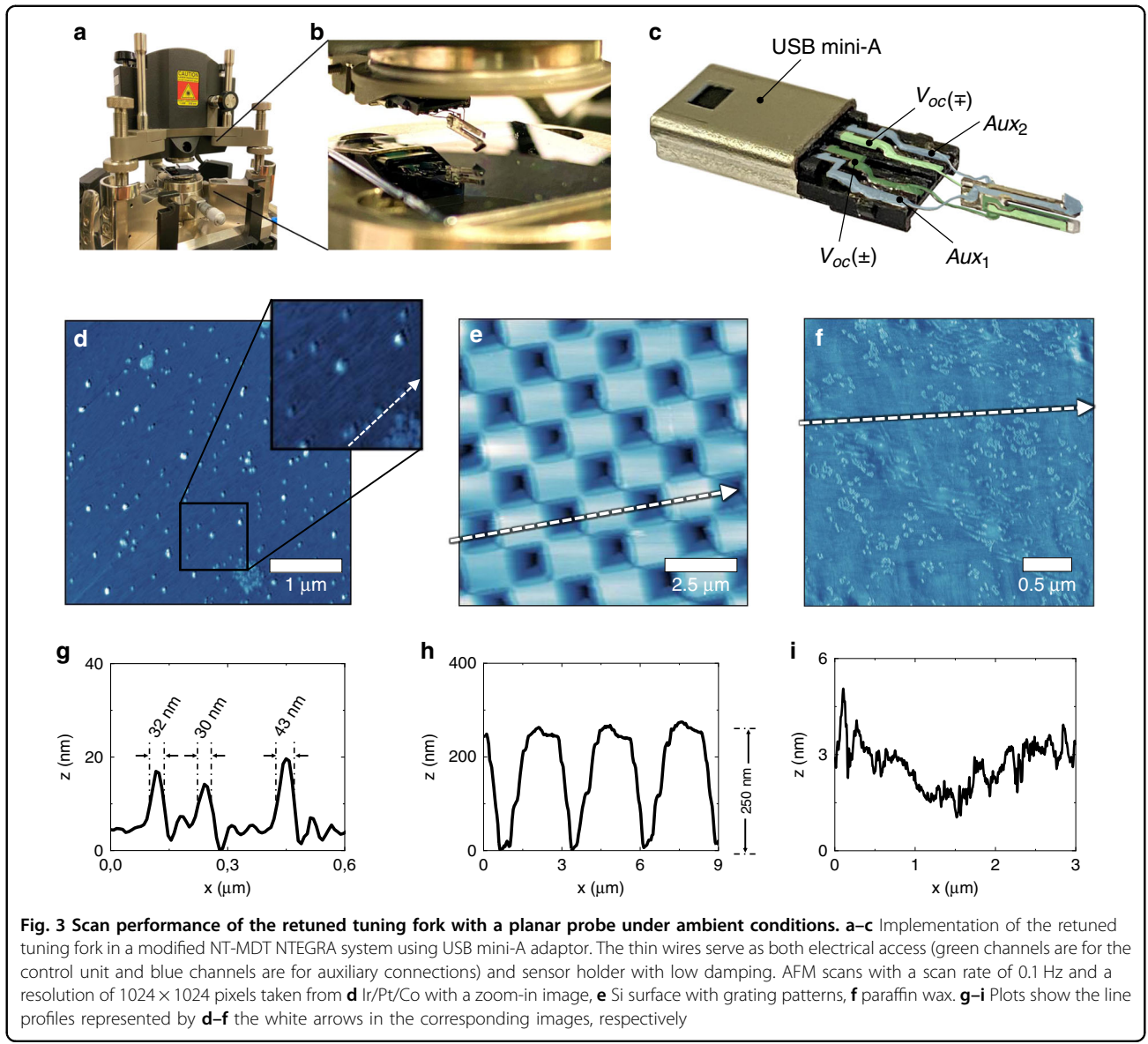

**Fig. 3 Scan performance of the retuned tuning fork with a planar probe under ambient conditions. a–c** Implementation of the retuned tuning fork in a modified NT-MDT NTEGRA system using USB mini-A adaptor. The thin wires serve as both electrical access (green channels are for the control unit and blue channels are for auxiliary connections) and sensor holder with low damping. AFM scans with a scan rate of 0.1 Hz and a resolution of 1024 × 1024 pixels taken from **d** Ir/Pt/Co with a zoom-in image, **e** Si surface with grating patterns, **f** paraffin wax. **g–i** Plots show the line profiles represented by **d–f** the white arrows in the corresponding images, respectively

The minimum detectable force ($F_{min}$) determines the sensitivity of an AFM[40]. Hida et al. expressed $F_{min}$ simply as[27],

$$F_{min} = \frac{d}{3A}\sqrt{\frac{4k_B TB}{2\pi}}\sqrt{\frac{k}{f_0 Q}}, \tag{1}$$

where $d$ is the tip-sample distance, $A$ is the oscillation amplitude, $k_B$ is the Boltzmann constant, $T$ is the environmental temperature, $B$ is the measurement bandwidth, $k$ is the spring constant, $f_0$ is the resonance frequency, and $Q$ is the quality factor. As shown in Eq. (1), the minimum sensing force depends on operation parameters, which are $d$ and $A$, and the oscillator's intrinsic variables $k$, $f_0$, and $Q$. As long as the amplitude never exceeds the distance to the surface, the ratio between $d$ and $A$ cannot be more than one $\left(\frac{d}{A} \leq 1\right)$ Therefore, minimizing $F_{min}$ would rely on optimization of the inherent characteristics of the oscillator via the parameters $k$, $f_0$, and $Q$.

Material properties and dimensional features of the oscillator are decisive for the parameters $k$ and $f_0$ as shown below,

$$k = \frac{Ebh^3}{4L^3}. \tag{2}$$

$$f_0 = \frac{1}{2\pi}\sqrt{\frac{k}{m}} = \frac{(1.875)^2}{4\pi}\sqrt{\frac{E}{3\rho}}\frac{h}{L^2} \tag{3}$$

where $E$ is Young's modulus, $b$, $h$, and $L$ are the width, thickness, and the length of one beam of the tuning fork,

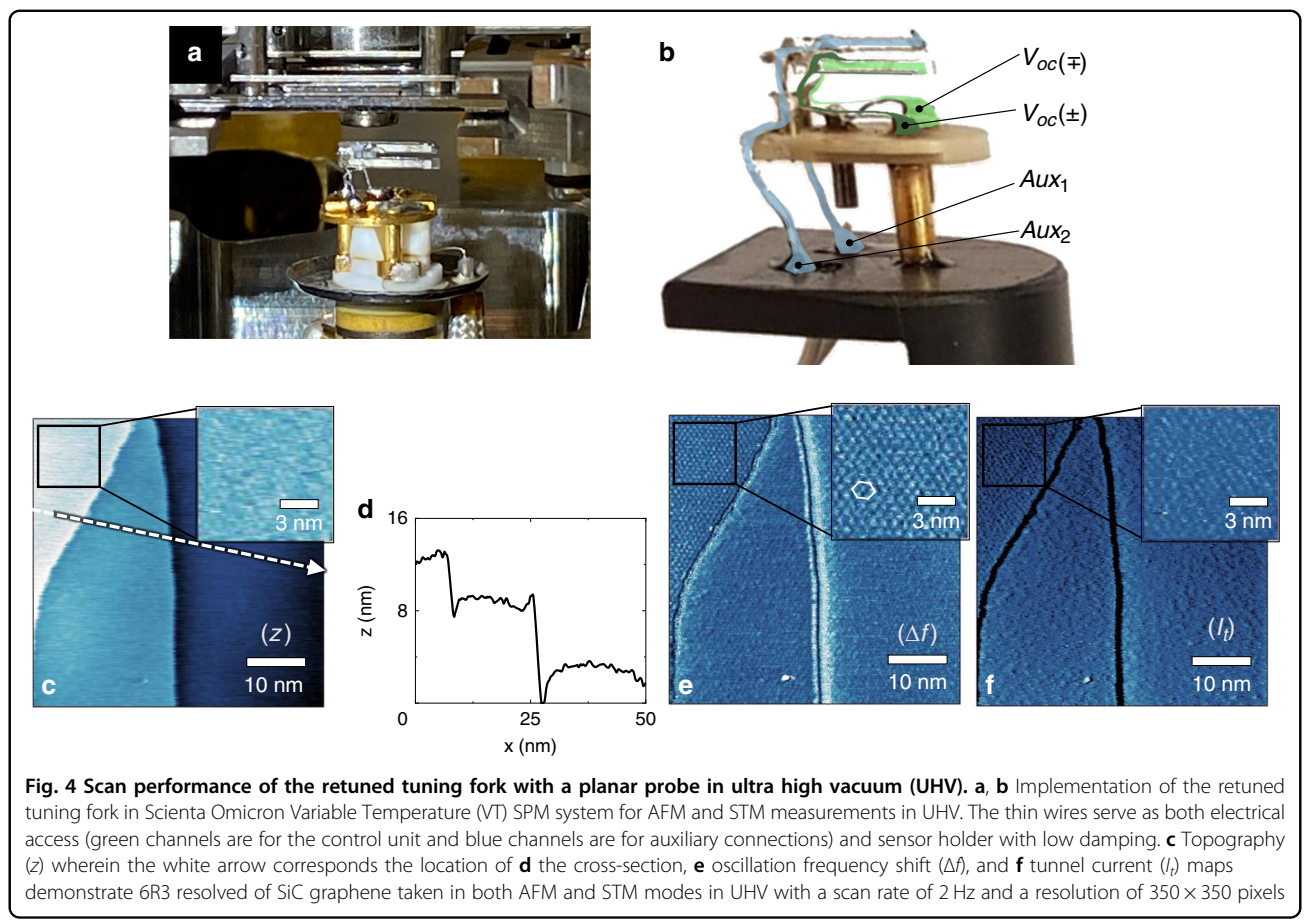

**Fig. 4 Scan performance of the retuned tuning fork with a planar probe in ultra high vacuum (UHV). a, b** Implementation of the retuned tuning fork in Scienta Omicron Variable Temperature (VT) SPM system for AFM and STM measurements in UHV. The thin wires serve as both electrical access (green channels are for the control unit and blue channels are for auxiliary connections) and sensor holder with low damping. **c** Topography (*z*) wherein the white arrow corresponds the location of **d** the cross-section, **e** oscillation frequency shift (Δ*f*), and **f** tunnel current (*I_t*) maps demonstrate 6R3 resolved of SiC graphene taken in both AFM and STM modes in UHV with a scan rate of 2 Hz and a resolution of 350 × 350 pixels

respectively, and *m* is the cantilever's mass with the density *ρ* of quartz[27].

The spring constant *k* of the quartz-tuning fork (AB38T) used in this study has the expressions *b*, *h*, and *L* optimized for carrying out oscillations in the short-range regime without jump-to-contact behavior. Within the intrinsic limits of the oscillator, we can only make minor alterations in these expressions *b*, *h*, and *L* to manipulate *k*. However, the resonance frequency *f₀* compensates for such a minor change in *k* considering their dependency for the same variables, as given in Eqs. (2) and (3). Therefore, the oscillation recovery of the designated oscillator seems possible only via optimizing the term *Q* to reduce *F*min.

The energy losses through several dissipation channels define the *Q* of an oscillator. For instance, a standard encapsulated AB38T model quartz-tuning fork shows *Q* value of $2 \times 10^4$ at its resonance of 32.768 kHz. However, this *Q* value significantly drops down to a few hundred when attaching an additional mass[27,41–43].

Improving environmental conditions can help reduce dissipation, such as placing the oscillator in a vacuum chamber or lowering the temperature. Unfortunately, this solution narrows down the scope of operations.

Another traditional option is minimizing the probe's size to lower the damping effect[27]. However, as previously explained, a tiny tip cannot provide sufficient flexibility in probe functionalization.

Therefore, we anticipate that only a structural reform in the quartz-tuning fork can pave the way for the applicational versatility of utilizing chip-like probes, maximizing the area for forming active and passive elements near the tip end. Learning from the benefits and limitations of the well-established qPlus, we sought alternative ways to compensate for such a misbalance to regain a high *Q* value after adding a chip-like probe. Eventually, we developed a method based on mass balancing in combination with electrode reconfiguration and floating-like holder-fixation, as discussed individually in the following.

**Mass retuning**

One of the ways to neutralize this consequence of attaching a probe is simply rebalancing the effective mass on the prongs by corresponding mass compensation[35]. Besides removing mass from the probe-carrying prong prior to tip attachment[34], as already introduced under Fabrication, one could also add a counter mass on the opposite tine[36]. However, adding a counter mass would create more complexity as it results in more deviation in the spring constant $k$ and the beam's length $L$. Considering that the resonances of both prongs will alter, it would be challenging to make this fine adjustment without knowing the reference frequency. Also, attaching a probe and its counter mass requires two times more glue, which potentially can cause more energy dissipation[44,45].

For these reasons, we have chosen the compensation method by removing the mass of the prong, as illustrated in Fig. 2. In this method, simple calculations allow to measure the mass of the planar prong and figure out how much material to polish away from the prong. As shown in Fig. 2d, removing quartz material with an ~45° angle allows the planar probe's optimal orientation and easy access for contact with fine adjustment. Since this process does not affect the second prong, it conserves the reference frequency for fine mass adjustment after probe fixation. Plots in Fig. 5 illustrate the oscillation characteristics near the resonance frequency before starting the fabrication process ($\Delta m = 0$), when removing some mass from the probe-carrying tine ($\Delta m > 0$), and after attaching an oversized planar probe ($\Delta m \approx 0$).

Figure 5 shows the changes in $Q$ and $f$ values during the fabrication process of one of the AB38T oscillators we used in our experiments. The dark-gray curve demonstrates that, after uncapping its capsule, the standard AB38T oscillator initially showed a $Q$ value around $1.2 \times 10^4$ when oscillating near 32.762 kHz under ambient conditions. After removing some mass and etching electrodes locally from the probe-carrying prong, the $Q$ dropped down to less than $1 \times 10^3$, and the resonance frequency increased to 35.397 kHz, as the red curve shows in Fig. 5. After retuning, the resonance frequency returned to 32.346 kHz, significantly closer to its initial value, and incremented the $Q$ value to $8 \times 10^3$, as the blue curve reveals in Fig. 5. This high $Q$ value reveals the efficiency of this retuning, especially for oversized probes.

### Electrode reconfiguration

The control over a specifically functionalized planar probe relies on an electrical link between the control unit and the "tip-on-chip" structure. Some advanced applications using complex on-chip probes can require multiple channels, but for simple ones, two auxiliary connections are sufficient. These external links are generally on chip-like probes connected via thin wires or fibers, such as SSFM[30], SHPM[31], and NFSMM[32,33]. However, these add-on wires modify effective mass and spring constant, which are difficult to compensate for, and an extra mechanical dissipation that drastically reduces the $Q$. Moreover, the long wires for current control introduce an additional capacitance affecting the readout signal.

Considering the complexities sourced by add-on components, we anticipated that the default electrode configuration surrounding the tuning fork might provide an alternative use besides only electrically tracing the oscillation. However, multi-purpose use of these electrodes can be tricky for the only oscillating prong of the qPlus systems. It is not easy to simultaneously oscillate a tip, read out the displacement, and carry separate signals between the probe and a control unit. For example, the two interconnections on the same cantilever would influence the electrical impedance resulting in poor oscillation by noisy transmission and inadequate control over the "tip-on-chip".

At this point, unlike the qPlus method, the two-free-prong approach can provide another unique solution: electrical insulation by the separation of two prongs. In principle, reading the tip deflection is possible via one prong in a mechanically driven oscillation within a very narrow spectrum near 32 kHz. So, with a minor structural alteration, we can assign the electrodes of the second prong for a different function, for example, carrying a current flow between the chip-like probe and the control unit. In this case, the independent circuits of current control do not affect the impedance of the circuit for the readout signal. This solution sacrifices half of the power of the sense signal since only half of the electrodes are in use, and therefore affects the ultimate sensibility. However, it resolves the problem of cross-talking between two signals and does the optimization of impedance that could have more severe consequences, as illustrated in Fig. 6.

Figure 6 shows the amplitude and phase characteristics of two similar retuned tuning-fork sensors having resonance frequencies close to each other. However, the first sensor has electrodes with their default electrode configuration, as shown in Fig. 6a. As the red curve shows, an additional capacitance exerts the phase back in the downward region towards −90° from the opposite phase around 90° when moving away from the resonance frequency[46]. The noisy

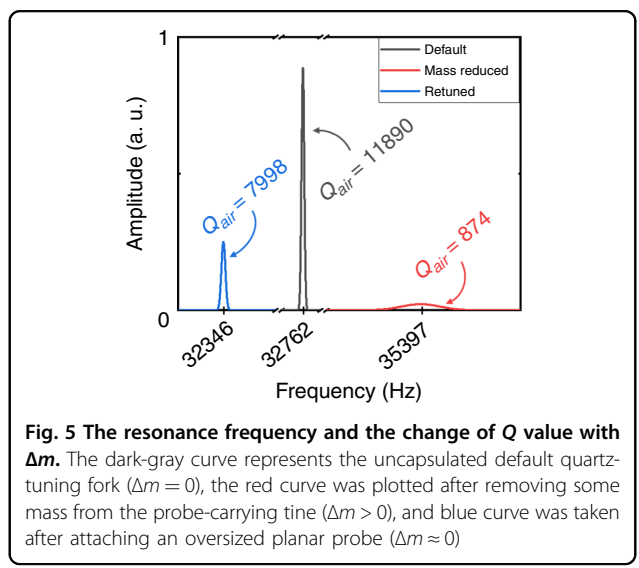

**Fig. 5 The resonance frequency and the change of $Q$ value with $\Delta m$.** The dark-gray curve represents the uncapsulated default quartz-tuning fork ($\Delta m = 0$), the red curve was plotted after removing some mass from the probe-carrying tine ($\Delta m > 0$), and blue curve was taken after attaching an oversized planar probe ($\Delta m \approx 0$)

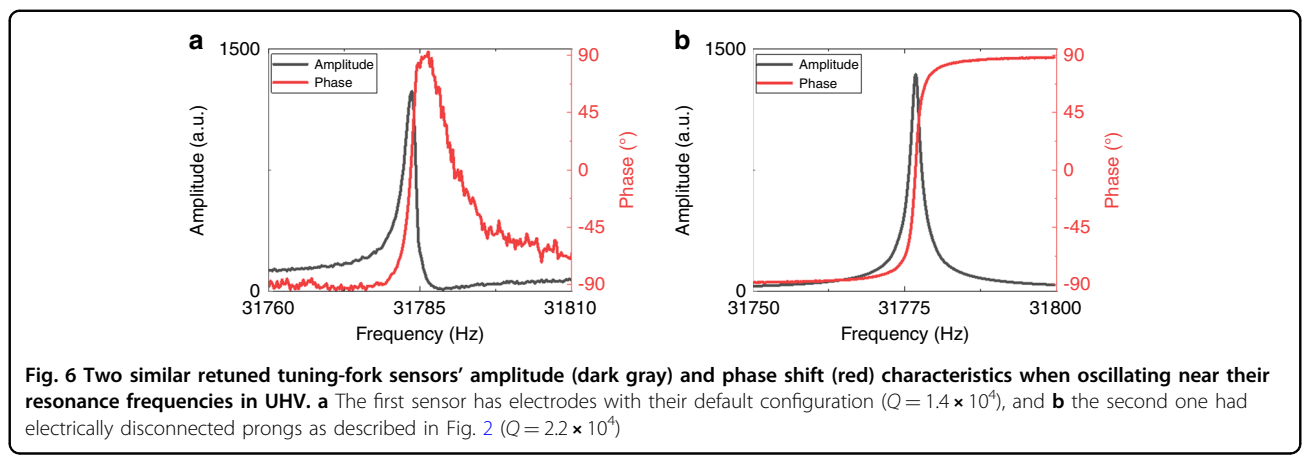

**Fig. 6 Two similar retuned tuning-fork sensors' amplitude (dark gray) and phase shift (red) characteristics when oscillating near their resonance frequencies in UHV. a** The first sensor has electrodes with their default configuration ($Q = 1.4 \times 10^4$), and **b** the second one had electrically disconnected prongs as described in Fig. 2 ($Q = 2.2 \times 10^4$)

background of the red plot might indicate a cross-talk or a picked-up external signal adding extra noise to the monitored readout. The asymmetric profile of the amplitude signal, shown in the dark-gray curve in Fig. 6a, confirms the perturbation of impedance via additive capacitance[46].

Next, we loaded the sensor with electrically disconnected prongs (Fig. 2c, d) for deducing its performance. As the plots in Fig. 6b demonstrate, several significant improvements stand out in the oscillation as symmetric amplitude characteristics, a much smoother phase shift curve, and an almost 60% higher $Q$ value ($2.2 \times 10^4$) than the default one ($1.4 \times 10^4$).

In conclusion, the results show how the reconfigured electrodes can optimize the performance by three significant achievements; supplying embedded electrodes preventing the system from add-on wire perturbation, canceling signals from stray capacitance and cross-talk, and improving the $Q$ remarkably.

## Floating-like holder-fixation

Mass balancing and electrode reconfiguration revealed a significant improvement in the $Q$ value, enabling the two-free-prong configuration to carry the oversized chip-like probes. Considering the difference between the $Q$ levels of the qPlus and the retuned tuning fork approaches when carrying a chip-like probe, also the fixation method seems to have a significant effect on the oscillation energy stored in the system. In order to draw a better picture of how fixation components potentially influence oscillators, we developed a COMSOL model showing the stress distribution of three systems oscillating at their resonance frequencies, as given in Fig. 7.

The first system is a standard quartz-tuning fork (AB38T) oscillator with no attached probe, as given in Fig. 7a. The fundamental mode of a standard tuning fork has a symmetric oscillation, as shown in Fig. 7a by the identically distributed high-stress regions around the knot point at the base where the two prongs meet. The area of no vibratory motion at the base, shown in dark blue, usually serves as a mounting spot of two rigid wires to transmit the readout signals without causing significant energy dissipation. We take the oscillation characteristics of the standard configuration as our optimal reference point for the following two variations.

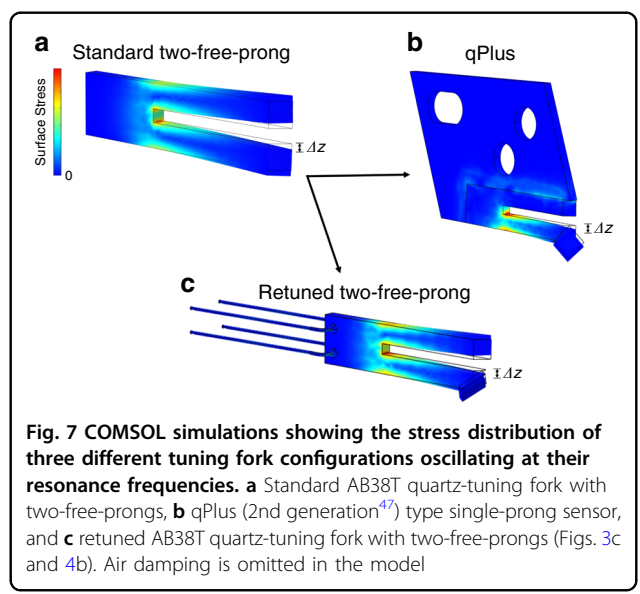

**Fig. 7 COMSOL simulations showing the stress distribution of three different tuning fork configurations oscillating at their resonance frequencies. a** Standard AB38T quartz-tuning fork with two-free-prongs, **b** qPlus (2nd generation[47]) type single-prong sensor, and **c** retuned AB38T quartz-tuning fork with two-free-prongs (Figs. 3c and 4b). Air damping is omitted in the model

However, any extra mass applied to only one tine naturally displaces the oscillation knot because of the strong coupling between two free prongs. A shifted $f$ and a drastic drop in $Q$ are inevitable consequences of such internally perturbed oscillation[41–43]. As a result, the standard configuration's sensitivity collapses when attaching an oversized probe, such as chip-like. Therefore, the qPlus, the second system shown in Fig. 7b, served as the only option for the tip-on-chips.

As given in Fig. 7b, the qPlus and its holder have vast contact surfaces surrounding the immobilized prong. A vibrating prong inherently attempts to transmit the oscillation energy to the other twin prong fixed to the holder. Because the second prong of the qPlus is immobilized, a considerable amount of the oscillation energy might transmit towards the holder components instead. Figure 7b demonstrates that the high-stress regions near the boundaries of the immobilized parts in the qPlus system might indicate an energy transmission towards the holder's contact surfaces, which might be a potential loss channel in the oscillation.

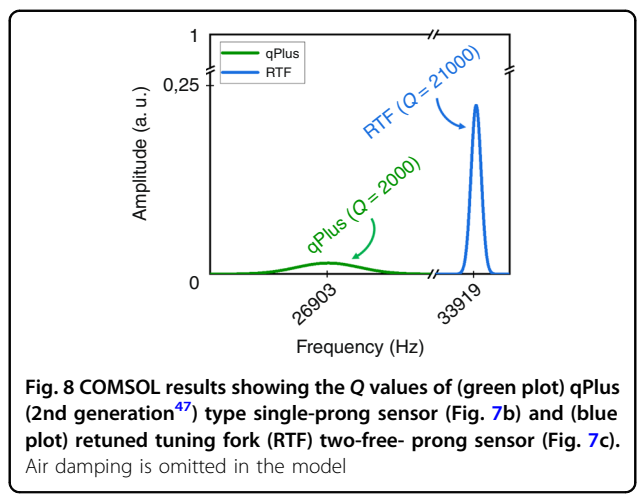

**Fig. 8 COMSOL results showing the *Q* values of** (green plot) qPlus (2nd generation[47]) type single-prong sensor (Fig. 7b) and (blue plot) retuned tuning fork (RTF) two-free- prong sensor (Fig. 7c). Air damping is omitted in the model

Such a holder fixation with vast contact surfaces is fundamentally not feasible for two- free-prong oscillators, which requires minimum external perturbation in the coupled oscillation beams. Therefore, our approach minimizes contact surfaces by employing soft wires used as holder components and signal transmitters, as demonstrated in Fig. 2d. Figure 7c presents the floating-like soft wire fixation for the retuned two-free-prong oscillator with a chip-like probe as the third and the last configuration. As given in Fig. 7c, the soft-wired fixation gives the retuned AB38T tuning fork a floating-like standing, and the stress distribution profile seems to be similar to its default characteristics shown in Fig. 7a. The fixation points where the soft wires link up with the oscillator are as far as possible from the high-stress regions, potentially resulting in less energy loss in the oscillation.

Figure 8 shows the COMSOL calculated *Q* and *f* values of the qPlus and the retuned tuning-fork oscillators given in Fig. 7b, c, respectively. As the plots in Fig. 8 reveal, the *Q* value of the retuned tuning fork fixed with soft wires is $2.1 \times 10^4$ at its resonance frequency of 33.919 kHz, which is $2 \times 10^3$ at its resonance frequency of 26.903 kHz for the qPlus. Thus, the calculated *Q* value of the retuned tuning fork is one order higher than the one of qPlus when carrying a chip-like probe. Moreover, there is also a big difference in their resonance frequencies. The retuned tuning fork's resonance frequency (33.919 kHz) is quite close to AB38T's default value (34.453 kHz), unlike the one of qPlus at 26.903 kHz.

As the COMSOL-based calculations and our experimental results correspondingly revealed, using soft wires as a floating-like fixation platform stands out as an efficient alternative that helps store the oscillating energy within the system. However, one might still worry about the possible eigenfrequencies of the wires that might overlap with the retuned tuning fork. According to the COMSOL model, long and straight pristine wires gave eigenfrequencies relatively near the spectrum of the tuning fork. However, this can easily be prevented by either shortening or randomly bending the wires. Shortened wires have shifted eigenfrequencies towards much higher values than our operational ranges, avoiding accidental coupling with the tuning fork. Besides, such a soft-wired fixation brings excellent flexibility for integrating with different systems, as demonstrated in our NT-MDT (Fig. 3a-c) and Omicron setups (Fig. 4a, b).

## Conclusion

In summary, our new approach of combining a retuned tuning-fork sensor with floating-like holder-fixation provides a strongly improved AFM performance. First of all, the retuning method provides broad flexibility for sensitivity recovery and compatibility with different systems in a broad range including traditional needle-like tips, cleaved wafers, and on-chip probes as long as a sharp edge serves as the scanning tip.

Combined with mass rebalancing, the reconfigured electrodes provide additional advantages, especially in reducing energy dissipation and electrical access to the tip-on-chip probes. Electrically disconnected electrodes on the upper and the lower prongs result in signal separation for the tip-on-chip control and the oscillation feedback. This configuration provides adequate electric access to the conductive probe with neither perturbation in the feedback signal nor additional wires causing extra damping. Hence, there is a significant improvement in the *Q* value.

As a result, we obtained AFM sensitivity with *Q* values reaching up to $1 \times 10^4$ in the air and $4 \times 10^4$ in UHV. This sensitivity level remains unmatched by the predecessor configurations carrying cleaved wafer tips or chip-like probes. The integration of the tip-on-chip with such a robust tuning fork drastically widens compatibility and sustainability, unlike the complex and fragile systems integrated on the conventional silicon cantilevers[18]. Thereby, using chip-like probes, the scan performance of our SPM approach revealed atomically resolved lateral resolution for STM, while almost a similar resolution was obtained for AFM as well.

## Outlook

Our approach to high sensitive tuning-fork utilization in AFM can pave the way for a new class of active chip-like probes with a size reaching up to $2 \times 2$ mm. The surface area of the attached probe reveals remarkable flexibility in implementing tip-on-chip designs in a broad range, from an integrated current-carrying wire to a complicated electronic circuit near the tip.

Unlike the conical or pyramidal geometry of the traditional AFM tips, the planar probes allow very efficient use of ultra-thin film engineering. High precision in the layer growth provides a broad material selection in a multilayer stacking and an unmatched tunability, for example, of an MFM-tip's stray field. Beyond proving a manifold but passive sensitivity, this film engineering on a vast and smooth surface of a planar probe lays a solid ground for micro-circuit tailoring as well.

Our pilot study[37] already demonstrated that a primitively structured circuit near the tip can generate a strong thermal gradient for thermomechanical patterning. Besides increasing the temperature, a current flow through a microwire can also serve as a heat-sensing element as a thermocouple or a thermistor.

In addition to thermal applications, the current-carrying microwire can also work as a magnetic element inducing Ørsted field in the tip vicinity. Such a locally induced field could manipulate spin textures in ferromagnetic thin films, for instance, labyrinth-like skyrmion patterns[48,49] or six-fold anisotropy of iron-garnets[50]. On the other hand, the induced field could also be used to orient the magnetization of a thin ferromagnetic coating at the apex of a planar probe. This type of control might provide a novel way of performing MFM with a switchable tip, which is more difficult in standard MFM tips.

Beyond a simple microwire, we can integrate much more complex electronic circuits for primary signal amplification and in situ signal procession. For instance, an in situ processors embedded near the tip can prevent the detected signal from traveling through long wires. Such local signal processing can allow working at higher frequencies, help reduce cross-talking and noise, and thus, simplify the signal transfer.

Flexibility in material selection envisions probe designs also for superconducting and photonic studies. For instance, a SQUID circuit or a superconducting bolometer near the tip can be used to monitor magnetic flux or infrared and millimeter waves, respectively. On the other hand, an implemented wave guide or a microstructured photonic crystal close to the tip can lead to Terahertz or optical range measurements in the near-field.

Considering its enhanced sensitivity, the widened scope of tip-on-chip design can convert the AFM from a trivial surface analysis tool with passive probes to a sophisticated device with active probes for complex characterization and nanomanipulation.

## Acknowledgements
This study is supported by NWO-TTW under project number 14715, by the French PIA under project number ANR-15-IDEX-04-LUE CAP-MAT, by the "FEDER-FSE Lorraine et Massif Vosges 2014–2020", and by the region Grand Est under the DESOMIN project. We thank Dr. Ivan Bykov and Dr. Stanislav Leesment from NT-MDT Spectrum Instruments for providing NTEGRA model AFM and technical support in the method development, Dr. Kıvanç Esat from Zurich Instruments for supplying MFLI model digital lock-in amplifier and assistance in the utilization, Wijnand Dijkstra for the contribution to the AFM sensor development.

## Conflict of interest
The authors declare no competing interests.